\documentclass[paper=letter]{scrartcl}
\usepackage[utf8]{inputenc}
\usepackage[english]{babel}
\usepackage{authblk}
\usepackage{siunitx}
\usepackage{amsmath}
\usepackage{amssymb}
\usepackage{graphicx}
\usepackage[includeheadfoot, headsep=.1in, margin={.6in,.6in}]{geometry}
\usepackage{tabularx}
\usepackage{hyperref}
\usepackage[margin=2cm, format=plain]{caption}
\usepackage[svgnames]{xcolor} 
\hypersetup{
    colorlinks=true,
    linkcolor=black,		
    urlcolor=DarkBlue,		
    anchorcolor = black,
    citecolor = DarkBlue,
    filecolor = black,
    breaklinks=false,}

\begin{document}

\title{Response to Comment on "Cell nuclei have lower refractive index and mass density than cytoplasm"}

\date{May 2, 2018}

\author{Paul M\"{u}ller and Jochen Guck\footnote{To whom correspondence should be addressed: jochen.guck@tu-dresden.de}}

\affil{Biotechnology Center of the TU Dresden, Germany}

\maketitle

\begin{center}
This text has been published in its final form at \url{https://doi.org/10.1002/jbio.201800095}.
\end{center}

\abstract{In a recent study entitled \textit{Cell nuclei have lower refractive index and mass density than cytoplasm}, we provided strong evidence indicating that the nuclear refractive index (RI) is lower than the RI of the cytoplasm for several cell lines (Schürmann, et al., J. Biophotonics, 2016).
In a complementary study, entitled \textit{Is the nuclear refractive index lower than cytoplasm? Validation of phase measurements and implications for light scattering technologies} (J. Biophotonics, 2017), Steelman et al. observed a lower nuclear RI also for other cell lines and ruled out methodological error sources such as phase wrapping and scattering effects.
Recently, M. A. Yurkin composed a comment on these two publications, entitled \textit{How a phase image of a cell with nucleus refractive index smaller than that of the cytoplasm should look like?}, putting into question the methods used for measuring the cellular and nuclear RI in the aforementioned publications by suggesting that a lower nuclear RI would produce a characteristic dip in the measured phase profile \textit{in situ}.
We point out the difficulty of identifying this dip in the presence of other cell organelles, noise, or blurring due to the imaging point spread function.
Furthermore, we mitigate Yurkin's concerns regarding the ability of the simple-transmission approximation to compare cellular and nuclear RI by analyzing a set of phase images with a novel, scattering-based approach.
We conclude that the absence of a characteristic dip in the measured phase profiles does not contradict the usage of the simple-transmission approximation for the determination of the average cellular or nuclear RI.
Our response can be regarded as an addition to the response by Z. A. Steelman, W. J. Eldridge, and A. Wax.
We kindly ask the reader to attend to their thorough ascertainment prior to reading our response.}

\section{Phase profile simulations}
\begin{figure}[h]
\begin{center}
\includegraphics[width=.8\linewidth]{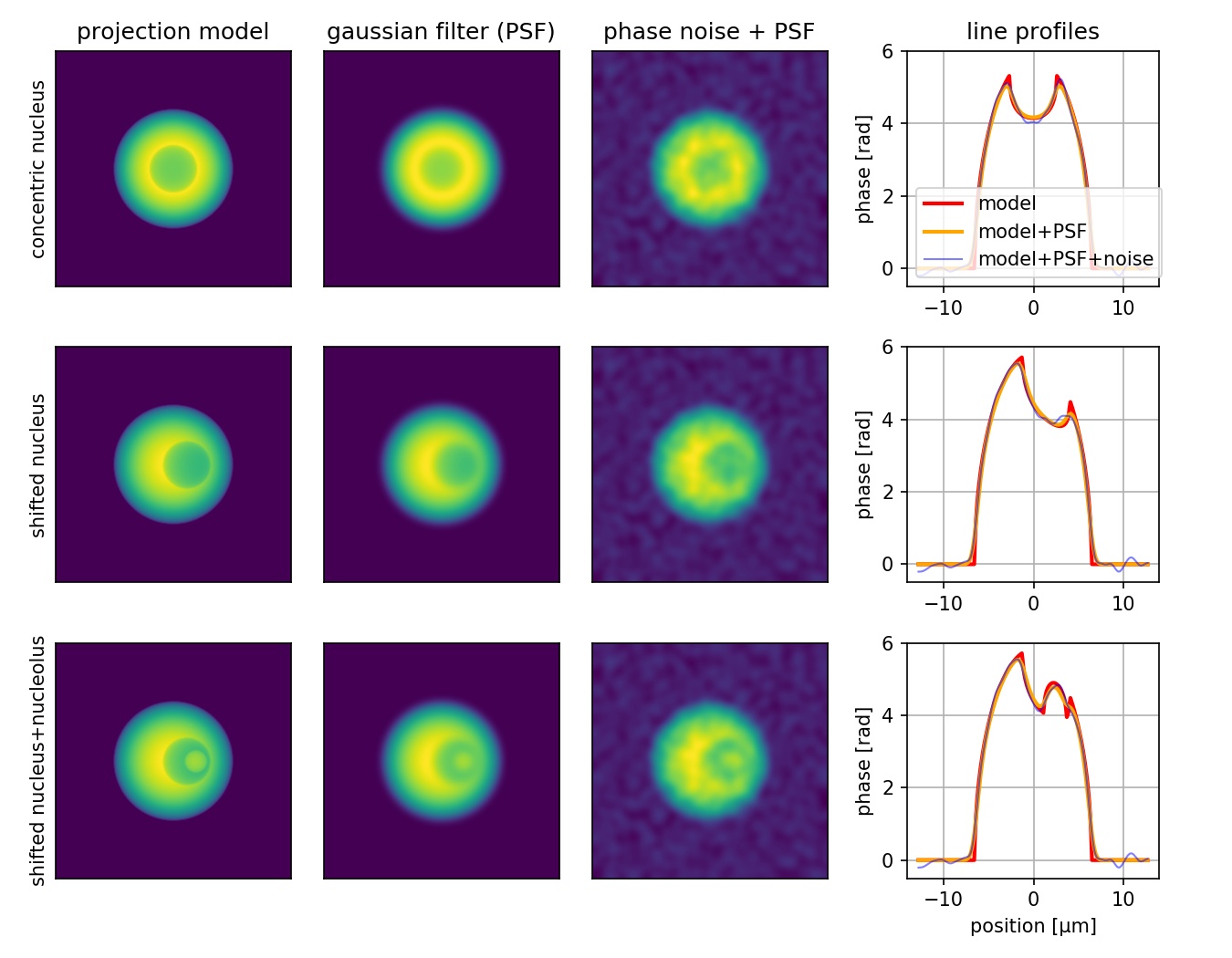}
\end{center}
\caption{Simple-transmission simulations. The phase images of a concentrically positioned nucleus (first row), a laterally shifted nucleus (second row), and a laterally shifted nucleus containing one nucleolus (third row) are shown. These are compared for the cases of no artifacts (first column), a convolution with the point spread function (PSF) with a gaussian filter using $\sigma=\SI{425}{nm}$ (second column), and the PSF combined with  additional phase noise (third column). The horizontal line profiles are shown in column four. The simulation parameters are identical to those in figure 2a of Yurkin's comment.\label{fig:sim}}
\end{figure}

The phase profile simulations presented by Yurkin \cite{Yurkin2018} oversimplify the image acquisition process and the imaged specimen. As discussed in the response by Steelman et al. \cite{Steelman2018}, several aspects are not taken into account. For instance, only nuclei positioned at the center of the cell are considered, whereas the nucleus may well be shifted laterally with respect to the cell center. Furthermore, the point spread function (PSF) must be taken into account. In reference \cite{Schuermann2016}, we specified a spatial resolution of \SI{850}{nm}. Therefore a convolution with a Gaussian of $\sigma=\SI{425}{nm}$ would be appropriate.
In addition, experimental phase data, especially when recorded with digital holographic microscopy (DHM), exhibits random phase noise that degrades the image quality.
Finally, inhomogeneities in the RI caused by the nucleolus or other cell organelles strongly perturb the phase image.
Figure~\ref{fig:sim} illustrates the contribution of these effects with a set of simulated phase images.
Thus, taking into account a few realistic conditions already reduces the visibility of characteristic dip feature severely.

\section{Phase image analysis}
In his comment, Yurkin concludes that the simple-transmission approximation does not correctly describe the scattering process in the case of two concentric spheres with RI values and sizes as found in \cite{Schuermann2016, Steelman2017}.
As noted by Steelman et al. in their response to Yurkin's comment, the phase-projection approximation is in fact known to yield accurate results for single-cell imaging.

\begin{figure}[h]
\begin{center}
\includegraphics[width=.5\linewidth]{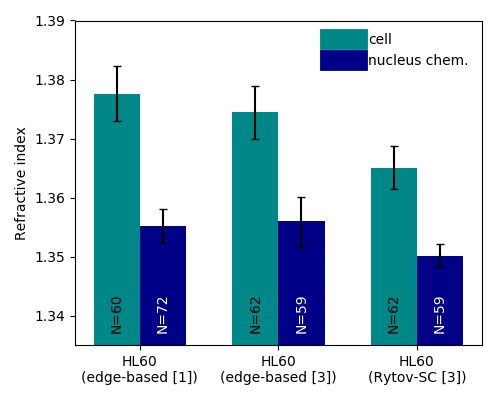}
\end{center}
\caption{Comparison of refractive index retrieval methods. The original results of reference \cite{Schuermann2016} are reproduced by our current analysis pipeline \cite{Mueller2018} with the edge-based method. The systematically-corrected Rytov approximation (see\cite{Mueller2018}), which is expected to yield more accurate results, only yields lower RI values while correctly reproducing the difference in RI between cells and isolated nuclei. For more information on the underlying data, please refer to figure 2 of reference \cite{Schuermann2016}.\label{fig:hl60}}
\end{figure}

Here, we would like to take the opportunity to point out a different issue with the phase image analysis employed in \cite{Schuermann2016}. The radii of the cells and nuclei were determined using an edge-detection (edge-based) approach. However, this approach underestimates the radius and thus overestimates the RI of the imaged object. Recently, we could show that this issue can be addressed by fitting a two-dimensional phase image with a corrected version of the Rytov approximation (Rytov-SC approach) to the experimental data \cite{Mueller2018}.
Figure \ref{fig:hl60} illustrates the difference between the edge-based and the Rytov-SC approaches by partly reproducing figure 2 of \cite{Schuermann2016}. Although the number of analyzed cells is different than in the original manuscript, both the edge-based and the Rytov-SC methods of our new analysis pipeline reproduce the difference between nuclear and cellular RI that was obtained previously.

\section{Continuing research}
To clarify the issue of the absence of the characteristic dip in the measured phase data, Yurkin proposed to employ rigorous light-scattering simulations with more sophisticated cell models. Indeed, it is known that large gradients in the RI distribution, such as those introduced by nucleoli, cannot be correctly resolved by both the simple-transmission approximation and the Rytov approximation (the RI of the nucleoli is slightly underestimated \cite{Mueller15}). However, as the RI difference between cell and nucleus is comparatively small, the simple-transmission approximation is sufficient to address the problem in question.

\begin{figure}[h]
\begin{center}
\includegraphics[width=.85\textwidth]{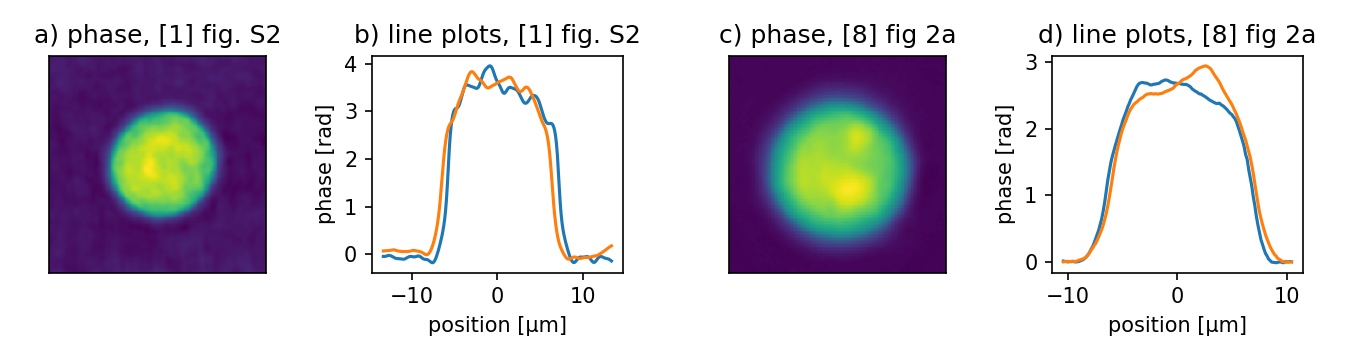}
\end{center}
\caption{Exemplary phase images of HL60 cells. a,b) Phase image and corresponding line plots through the image center for the HL60 cell shown in reference \cite{Schuermann2016}, supplementary figure S2. c,d) Phase image and corresponding line plots through the image center for the HL60 cell shown in reference \cite{Schuermann2017}, figures 2 and 4.\label{fig:data}}
\end{figure}

As pointed out by Steelman et al., optical diffraction tomography (ODT) is a suitable method to measure the RI difference between nucleus and cytoplasm, as there are already highly suggestive ODT studies pointing at a lower nuclear RI for HT29 cells (\cite{Kim2014}, fig.  and \cite{Sung2009}, fig. 6), HaCaT and SG cell lines and cancerous CA9-22 and BCC cell lines (\cite{Kim2016}, fig. 8), or HL60 cells (\cite{Schuermann2017}, fig. 4).
Figure \ref{fig:data} helps to illustrate the necessity of ODT over an analysis based on a single phase image. Both cells shown do not exhibit a characteristic dip in the phase profile. Figure \ref{fig:data}a only shows a barely visible ring-like structure at the cell center. In figure \ref{fig:data}c, no ring-like structure is visible. Nevertheless, an ODT analysis of the cell shown in figure \ref{fig:data}c revealed a nuclear RI that is lower than that of the cytoplasm using 3D refractive index and fluorescence colocalization tomography (\cite{Schuermann2017}, fig. 4). Thus, to determine the nuclear RI, a rigorous numerical approach is not necessary.

\section{Conclusion}
We showed that experimental noise, the imaging point spread function, and natural inhomogeneities within the imaged cell lead to distortions that make it difficult, if not impossible, to draw conclusions regarding the position, size, or RI of the nucleus \textit{in situ}.
In addition, we emphasized the fact that the simple-transmission approximation is indeed a valid tool for the relative comparison of nuclear and cellular RI, although more accurate tools have recently been developed.
Furthermore, we pointed out that a tomographic approach, such as ODT, has already been proven to resolve intracellular structures and that there is strong evidence in literature that suggests a lower nuclear RI for several cell lines.
Thus, we believe that a rigorous treatment of light scattering, as proposed by Yurkin, is not necessary to address the question whether the nuclear RI is lower than that of the cytoplasm. The fact that we do not observe the characteristic dip in the phase images, as predicted by Yurkin, does not contradict our analysis method but rather indicates an oversimplified model.

\bibliographystyle{ieeetr}
\bibliography{response}

\end{document}